\documentclass[aps,twocolumn,prl,tightenlines,floatfix,showpacs]{revtex4}
\usepackage[dvips]{graphicx}
\usepackage[english]{babel}  
\usepackage{amsmath}  
\usepackage{amssymb}  
\usepackage{slashed}  
\usepackage{feynmf}
\usepackage{marginnote}

\begin{document}

\title{Theory of Diamagnetism in the Pseudogap Phase: Implications from
the
Self energy of Angle Resolved Photoemission}

\author{Dan Wulin and K. Levin}

\affiliation{James Franck Institute and Department of Physics,
University of Chicago, Chicago, Illinois 60637, USA}

\date{\today}
\pacs{}

\begin{abstract} 
In this paper we apply the emerging- consensus understanding of
the fermionic self energy deduced from angle resolved
photoemisssion spectroscopy (ARPES) experiments to deduce the
implications for orbital diamagnetism in the underdoped cuprates.
Many theories using many different starting points have
arrived at a broadened BCS-like form for the normal state
self energy associated with a $d$-wave excitation
gap, as is compatible with ARPES data. Establishing compatibility
with the f-sum rules, we show how
this self energy, 
along with the constraint that
there is no Meissner effect in the normal phase 
are sufficient to deduce
the orbital susceptibility.
We conclude, moreover, that diamagnetism is large for
a $d$-wave pseudogap. Our results should apply rather widely
to many theories of the pseudogap, independent of the microscopic details.
\end{abstract}

\maketitle

\section{Introduction}

The origin of the pseudogap in high temperature superconductors
is still very much under debate.
While there are indications of an alternative order parameter
\cite{Taillefer4,Hinkov}
it is widely accepted that the diamagnetism which is associated
with the pseudogap is indicative of some form of
precursor pairing. 
Nevertheless, even within the various schools which posit 
preformed pairs there is controversy about their
nature and origin. What would be particularly useful,
given this multitude of scenarios is to build an understanding
of the pseudogap phase based on aspects
of phenomenological (and microscopic) theories about which
there is some degree of convergence.

In this paper we focus on the diamagnetism of the cuprates
(in the linear response
regime) and show how it is connected to an emerging- consensus
understanding \cite{Malypapers,Normanphenom,SenthilLee,Chubukov2} of
angle resolved photoemission experiments. These experiments
have demonstrated that a simple broadened BCS form for the
self energy in the presence of a pseudogap
$\Delta_{pg,\bf k}$
\begin{eqnarray}
\Sigma_{pg,K}=-i\gamma+\frac{\Delta_{pg,\bf k}^2}{i\omega_n+\xi_{\bf k}+i\gamma}
\label{eq:1}
\end{eqnarray}
works rather well.
Indeed, this form of the self energy is widely subscribed to
\cite{Malypapers,Normanphenom,SenthilLee,Chubukov2}
in diverse theories of the pseudogap. Here we exploit
this equation (and the related fermiology constraining the
bandstructure
$\xi_{\bf k}$) 
in conjunction
with the important constraint that there is
no Meissner effect in the pseudogap phase, to arrive at a form
for the orbital susceptibility.
This relation between transport and self energy is well
known from the Ward identities.
In a closely related fashion we justify our transport expressions
for the diamagnetic response
by showing that the related current-current correlation function
analytically satisfies the transverse and longitudinal f-sum rules.
In this sense this understanding of diamagnetism should have
a generality which goes beyond a particular microscopic or
phenomenological approach.

Recent experiments measuring the orbital susceptibility in a 
variety of cuprate superconductors report anomalously large diamagnetism 
\cite{Ong_Dia,Ong2_Dia}). 
This diamagnetism onset is well correlated with
the onset of an enhanced  
Nernst signal above $T_c$ in the same materials \cite{Wang_Dia}.
It is not likely that low dimensional, critical fluctuations are
responsible, because
the diamagnetism persists beyond the expected critical regime.
Moreover, there are reports of strong non-linear effects, although
they will not be the topic of this paper.

This pseudogap phase has been very systematically studied
in angle resolved photoemission spectroscopy (ARPES) experiments 
which lead to the phenomenological expression in Eq.~(\ref{eq:1})
as well as the underlying fermiology.
An important feature in the precursor superconductivity
approaches is the
Fermi
``arcs", (as distinguished from Fermi ``pockets"). 
These are understood to arise from
the phenomenological broadening factor $\gamma$ in Eq.~(\ref{eq:1})
which
leads to a smearing out of the $d$-wave nodes.
Also relevant are the collapse of the arcs below
$T_c$ and the related ``two gap" phenomena which
appear in the superconducting phase. Importantly, they follow 
naturally \cite{FermiArcs,ourarpes} from
Eq.~(\ref{eq:1}) if one adds to $\Sigma_{pg}$ a co-existing conventional
condensate contribution $\Sigma_{sc}$,
(of the same form but with the
order parameter contribution, $\Delta_{sc}^2$
appearing and the associated $\gamma= 0$). 
Theoretical approaches to 
diamagnetism have not incorporated this ARPES constraint,
largely because they are based on thermodynamic rather than
transport schemes. 
Moreover, for the most part the emphasis has been on the contribution
of bosonic pairs either in fluctuation 
\cite{LarkinVarlamov,Huse_Dia,Huse2_Dia}
or real space pairing
schemes.
\cite{Alexandrov_Dia}.
Nevertheless, physically, one might expect the total diamagnetic
response to reflect fermions as well as these pair correlations.
Indeed, this has to be crucial when the fermionic self energy
is incorporated and sum rules are addressed.
Alternative theories
based on vortex liquids \cite{Anderson_Dia} 
or d-density wave states \cite{Sau_Dia}
have addressed complementary physics.

\begin{figure*}
\begin{center}
\includegraphics[width=5.5in,clip=true]{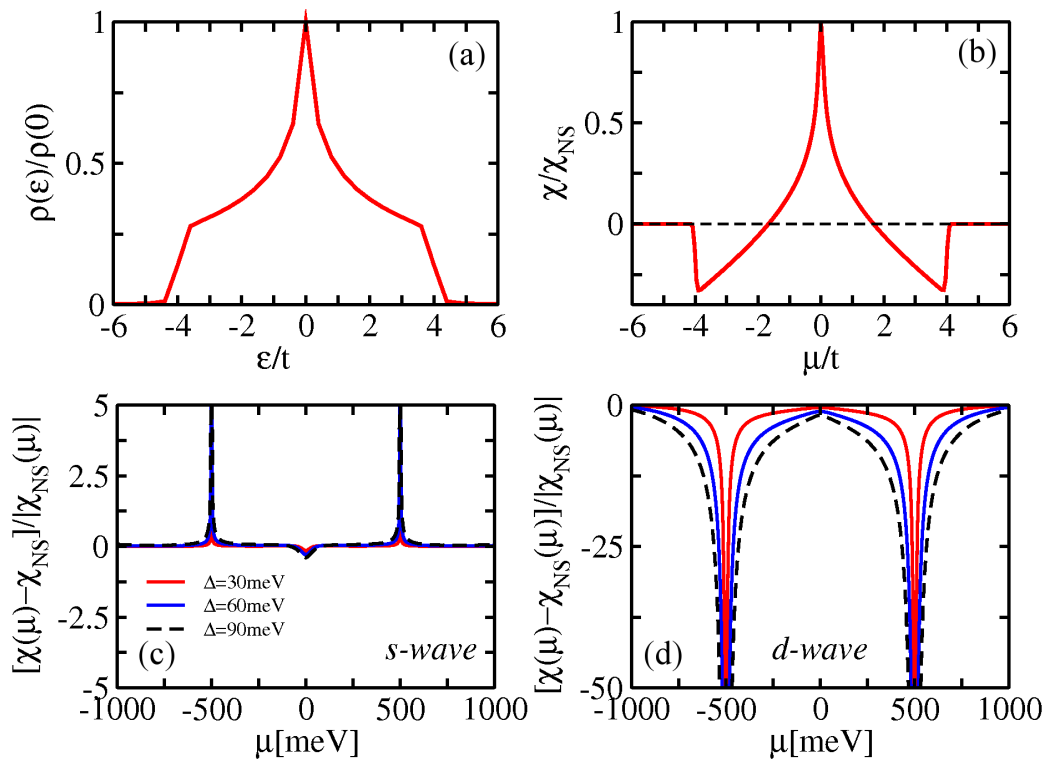}
\caption{Summary Figure showing the effect of the pseudogap for
$s$ and $d$-wave lattices via plots as functions of
band filling or chemical potential, $\mu$. (a) corresponds to the assumed
density of states, (b) to the normal state susceptibility
while (c) and (d) illustrate the
the difference between the pseudogap and normal state
$\chi^{\textrm{dia}}$ normalized by the absolute value of $\chi^{\textrm{dia
}}$ in the normal state at the same $\mu$. In (c) the pseudogap is taken to have an $s$-wave form fac
tor. (d) The analogous plot but for a $d$-wave order parameter.}
\label{fig:chi_NLattice_Dia}
\end{center}
\end{figure*}

\subsection{Overview of Theory and Results}

It is useful to begin with an overview of the connection
between ARPES and diamagnetism. The
reasoning is relatively simple and is best communicated by
presenting the theory first in summary form.

For the sake of notational simplicity, the equations
we present here are for the negligibly small $\gamma$ limit, although
the general line of reasoning can readily accomodate any
size $\gamma$, as discussed in the
Appendix. The transport-based expression for the diamagnetic response is
\begin{equation}
\chi^{\textrm{dia}}=-\displaystyle{\lim_{q_y\rightarrow0}}\textrm{Re}\Bigg[\frac{
P_{xx}(\textbf{q},\omega=0)+(n/m)_{xx}}{\textbf{q}^2}
\Bigg]_{q_x=q_z=0}\label{eq:chi_Dia}
\end{equation}
Here
$\chi^{\textrm{dia}}$ is written in terms of the current-current correlation function $\overleftrightarrow{P}(\textbf{q},\omega)$ and the diamagnetic current tensor $\overleftrightarrow{n}/m$. 
The absence of a Meissner effect 
is equivalent to the vanishing of the electromagnetic response kernel above $T_c$ at zero momentum and frequency, i.e. the response kernel must satisfy
\begin{eqnarray}
\overleftrightarrow{P}(0,0)+\frac{\overleftrightarrow{n}}{m}=0,T\geq T_c\label{eq:meis_Dia}
\end{eqnarray}
When Eq.\ref{eq:meis_Dia} is satisfied, the first nonzero term in the numerator of Eq.\ref{eq:chi_Dia}  comes in at $\mathcal{O}(\textbf{q}^2)$, rendering $\chi^{\textrm{dia}}$ well-behaved. The analogue statement below $T_c$ is that the right hand side of Eq.\ref{eq:meis_Dia} equals $\overleftrightarrow{n}_s/m$, the superfluid density tensor.

It is helpful to recast the expression for $\overleftrightarrow{n}/m$ in a way that closely mirrors the structure of a response function:
$\frac{\overleftrightarrow{n}}{m}=2\displaystyle{\sum_K}\frac{\partial^2\xi_{\bf k}}{\partial\bf k\partial\bf k}G_K$ can be rewritten after 
some straightforward algebra as
\begin{eqnarray}
\overleftrightarrow{P}(0,0)=2\displaystyle{\sum_{K}}\frac{\partial\xi_{\bf k}}{\partial\bf k}\frac{\partial\xi_{\bf k}}{\partial\bf k}G_K^2\Big(1-\Delta_{pg,\bf k}^2G^2_{0,-K}\Big)\label{eq:P0_Dia}
\end{eqnarray}

Now let us consider the relation to Landau diamagntism.
Ignoring lattice effects for the moment,
the standard expression for the orbital susceptibility is of the form (\cite{Ashcroft_Dia,Schmid_Dia})
\begin{eqnarray}
\chi^{\textrm{dia}}\propto\frac{-n}{m} \frac{e^2}{c^2}\langle\textbf{r}^2\rangle
\label{eq:chiGL_Dia}
\end{eqnarray}
where $n$ and $m$ represent the density and mass of the charge
carriers in Landau orbits. The source of large
conventional fluctuation diamagnetism \cite{LarkinVarlamov}
is the characteristic size 
$\langle\textbf{r}^2\rangle$
which is presumed equal to the correlation length and
which diverges
as the transition is approached. Consequently,
even in the absence of fluctuations,
one might expect that $s$-wave pairs would have significantly less
diamagnetism than the more extended $d$-wave pairs 
of the cuprates, as we demonstrate here.

We will show that
the orbital susceptibility deriving from Eq.~(\ref{eq:1}) arises from
both fermionic and bosonic contributions.
Diamagnetism comes predominantly from the latter and one can
understand this physically as associated with the 
general enhancement of bosonic
contributions to transport in the vicinity of Bose
condensation. 
A strongly
peaked Bose distribution function (in momentum space)
readily accomodates
a redistribution of particles leading to large
transport responses even in the presence of weak perturbing
fields. Indeed, this is the origin of superconducting fluctuation
contributions, in general. 
This is in contrast to
fermionic transport, which is
restricted by the Pauli principle.

We note that 
\cite{Vignale_Dia}
even in a zero
gap normal state, bandstructure effects
can yield a paramagnetic orbital susceptibility, particularly
near Van Hove singularities. Thus 
one might expect the net diamagnetic contribution will
be largest away from the Van Hove points. An additional
effect of the fermionic contribution derives from the fact that
an
excitation gap will reduce the number of available fermions $n$.
Because of nodal fermions,
$d$-wave pairing in the pseudogap phase is expected to
yield more diamagnetism.
In this paper we will show how all of these effects combine
to yield a rather large diamagnetic response strongly associated with
a $d$-wave pseudogap.

We present in this introduction an important inference which
will be discussed in more detail later and which
allows us to arrive at 
an extension of Eq.\ref{eq:P0_Dia} to finite momentum and frequency in the
form
\begin{eqnarray}\label{eq:P_Dia}
{\overleftrightarrow{P}}(\textbf{q},i\Omega_m)&=& {2}\displaystyle{\sum_K}\frac{\partial\xi_{\textbf{k}+\textbf{q}/2}}
{\partial\textbf{k}}\frac{\partial\xi_{\textbf{k}+\textbf{q}/2}}
{\partial\textbf{k}}\Big[G_KG_{K+Q}
\\&&-\Delta_{pg,\textbf{k}}\Delta_{pg,\textbf{k}+\textbf{q}}G_{0,-K-Q}G_{0,-K}
G_{K+Q}G_K\Big] \nonumber
\end{eqnarray}
One could anticipate such an answer
using
the standard normal state expression for the current-current correlation function (c.f. \cite{Fetter_Dia})
or alternatively the counterpart for the standard BCS current-current
correlation function (where one has to be careful to
enforce a Meissner effect, rather than its absence in
the pseudogap phase). But the strongest support for
Eq.~(\ref{eq:P_Dia}) is the demonstration that it analytically
satisfies the f-sum rules. 
These are discussed in more detail in Appendix A.
Importantly
Eq.~(\ref{eq:P_Dia}) 
provides the input one needs to arrive at transport properties
(including the complex conductivity) that arise from Ward identity
compatibility with the ARPES-derived self energy.

It is useful to end this introduction with a summary figure
addressing the implications of the pseudogap as it appears in
Eq.~(\ref{eq:1}) on the orbital susceptibility.
The four panels in this figure correspond to (a) the
assumed density of states as a function of
energy and (b) the counterpart normal
state orbital susceptibility as a function of band
filling. Note that the simple Landau diamagnetism of
jellium can give rise to paramagnetism in a tight binding
band, particularly near Van Hove singularities.
In (c) we indicate the \textit{change} in the orbital
susceptibility associated with the pseudogap as a function
of band filling for $s$-wave pairing. The counterpart
figure for the $d$-wave case is shown in (d).
A comparison shows that the $d$-wave pseudogap is
associated with substantially enhanced diamagnetism.
This figure will be discussed in more detail in the context of
our numerical results.

\section{Theory of Diamagnetic Susceptibility}

We begin by rewriting
the expression for $\overleftrightarrow{n}/m$:
\begin{eqnarray}
=-2\displaystyle{\sum_K}\frac{\partial\xi_{\bf k}}{\partial\bf k}\frac{\partial G_K}{\partial\bf k}\label{eq:int1_Dia}
\end{eqnarray}
following an integration by parts. 
Next we rewrite the derivative of the Green's function using the identity $\partial G_K/\partial\textbf{k}=-G^2_K\partial G^{-1}_K/\partial\bf k$, where the derivative of the inverse Green's function is straightforward to evaluate in terms of the ARPES-derived $\Sigma_{pg,K}$,
\begin{eqnarray}
\frac{\partial G^{-1}_{\bf k}}{\partial \bf k}=\frac{\partial G_{0,K}^{-1}}{\partial\bf k}-\frac{\partial\Sigma_{pg,K}}{\partial\bf k}=-\frac{\partial\xi_{\bf k}}{\partial\bf k}-\frac{\partial\Sigma_{pg,K}}{\partial\bf k}\label{eq:int2_Dia}
\end{eqnarray}
From Eq.\ref{eq:meis_Dia}, the current-current correlation function at $\textbf{q},i\Omega=0$  necessarily 
yields Eq.\ref{eq:P0_Dia}.

The structure of Eq.\ref{eq:P0_Dia} is intriguing: the first term in parenthesis is the usual electromagnetic response of fermionic quasiparticles, while the second term arises from the presence of  pseudogap correlations 
and has the appearance of being a correction to the bare electromagnetic vertex. 

We then posit 
Eq.\ref{eq:P_Dia} as
the natural extension to arbitrary four-vector
$Q$, and this is supported by the transverse and longitudinal f-sum rules.
For the latter
we have to prove 
the longitudinal and transverse f-sum rules. Previously \cite{OurAC,LongConduct}
we have focused on the second of these, which will be summarized
here in the Appendix. We now address the longitudinal sum rule

\begin{eqnarray}\label{sr}
\int^{+\infty}_{-\infty}\frac{d\omega}{\pi}\big(-\frac{\textrm{Im}\chi_{JJ}^L(\omega,\mathbf{q})}{\omega}\big)=\frac{n}{m},
\end{eqnarray}
We define
$f^{\pm}\equiv f(E_{\bf k})$ as the fermi function, $E^{\pm}\equiv=E_{\textbf{k}\pm\text
bf{q}/2}$, $\xi^{\pm}\equiv\xi_{\textbf{k}+\textbf{q}/2}$, and $\Delta_{pg}^{\pm}\equiv\Delta_
{pg,\textbf{k}\pm\textbf{q}/2}$.
The normal state fermions have dispersion
$E_{\mathbf{k}} \equiv \sqrt{ \xi_{\mathbf{k}}^2 + \Delta_{pg}^2(T)  }$. 
The longitudinal component of $\tensor{\chi}_{JJ}$ is defined as $\chi_{JJ}^L\equiv\hat{\mathbf{q}}\cdot\tensor{\chi}_{JJ}\cdot\hat{\mathbf{q}}=\mathbf{q}\cdot\tensor{\chi}_{JJ}\cdot\mathbf{q}/q^2$.
We need several simple relations
\begin{eqnarray}
& &\textrm{Im}\frac{1}{\omega\pm(E^+_{\mathbf{p}}\pm E^-_{\mathbf{p}})+i\delta}=-\pi\delta\big(\omega\pm(E^+_{\mathbf{p}}\pm E^-_{\mathbf{p}})\big),\nonumber\\
& &\frac{\mathbf{p}\cdot\mathbf{q}}{m}=\xi^+_{\mathbf{p}}-\xi^-_{\mathbf{p}}
\end{eqnarray}
along with
\begin{eqnarray}
& &(\xi^+_{\mathbf{p}}-\xi^-_{\mathbf{p}})(E^+_{\mathbf{p}}E^-_{\mathbf{p}}-\xi^+_{\mathbf{p}}\xi^-_{\mathbf{p}}+\Delta^2_{\textrm{pg}})  
\\
&=&(\xi^+_{\mathbf{p}}E^-_{\mathbf{p}}-\xi^-_{\mathbf{p}}E^+_{\mathbf{p}})(E^+_{\mathbf{p}}+E^-_{\mathbf{p}}), ~~~~~~~\rm{and}~~~~,\nonumber \\
& &(\xi^+_{\mathbf{p}}-\xi^-_{\mathbf{p}})(E^+_{\mathbf{p}}E^-_{\mathbf{p}}+\xi^+_{\mathbf{p}}\xi^-_{\mathbf{p}}-\Delta^2_{\textrm{pg}})
\nonumber \\
&=&(\xi^+_{\mathbf{p}}E^-_{\mathbf{p}}+\xi^-_{\mathbf{p}}E^+_{\mathbf{p}})(E^+_{\mathbf{p}}-E^-_{\mathbf{p}}).
\end{eqnarray}

We have
\begin{widetext}
\begin{eqnarray}\label{E2}
& &\int^{+\infty}_{-\infty}\frac{d\omega}{\pi}\big(-\frac{\textrm{Im}\chi_{JJ}^L(\omega,\mathbf{q})}{\omega}\big) \nonumber \\
&=&\sum_{\mathbf{p}}\frac{\mathbf{p}\cdot\mathbf{q}(\xi^+_{\mathbf{p}}-\xi^-_{\mathbf{p}})}
{mq^2}\Big\{\frac{E^+_{\mathbf{p}}E^-_{\mathbf{p}}-\xi^+_{\mathbf{p}}\xi^-_{\mathbf{p}}+\Delta^2_{\textrm{pg}}}{E^+_{\mathbf{p}}E^-_{\mathbf{p}}}\frac{1}{E^+_{\mathbf{p}}+E^-_{\mathbf{p}}}\big[1-f(E^+_{\mathbf{p}})-f(E^-_{\mathbf{p}})\big] \nonumber \\
& &\qquad\qquad\qquad\qquad-\frac{E^+_{\mathbf{p}}E^-_{\mathbf{p}}+\xi^+_{\mathbf{p}}\xi^-_{\mathbf{p}}-\Delta^2_{\textrm{pg}}}{E^+_{\mathbf{p}}E^-_{\mathbf{p}}}\frac{1}{E^+_{\mathbf{p}}-E^-_{\mathbf{p}}}\big[f(E^+_{\mathbf{p}})-f(E^-_{\mathbf{p}})\big]\Big\},
\end{eqnarray}

This yields
\begin{eqnarray}\label{E2}
& &\int^{+\infty}_{-\infty}\frac{d\omega}{\pi}\big(-\frac{\textrm{Im}\chi_{JJ}^L(\omega,\mathbf{q})}{\omega}\big)\nonumber\\
&=&\sum_{\mathbf{p}}\frac{\mathbf{p}\cdot\mathbf{q}}{mq^2}\big[\frac{\xi^+_{\mathbf{p}}}{E^+_{\mathbf{p}}}\big(1-2f(E^+_{\mathbf{p}})\big)-\frac{\xi^-_{\mathbf{p}}}{E^-_{\mathbf{p}}}\big(1-2f(E^-_{\mathbf{p}})\big)\big]\nonumber\\
&=&\sum_{\mathbf{p}}\frac{\mathbf{p}\cdot\mathbf{q}}{mq^2}\big[1-\frac{\xi^-_{\mathbf{p}}}{E^-_{\mathbf{p}}}\big(1-2f(E^-_{\mathbf{p}})\big)\big]-\sum_{\mathbf{p}}\frac{\mathbf{p}\cdot\mathbf{q}}{mq^2}\big[1-\frac{\xi^+_{\mathbf{p}}}{E^+_{\mathbf{p}}}\big(1-2f(E^+_{\mathbf{p}})\big)\big].
\end{eqnarray}
\end{widetext}
Changing variables $\mathbf{p}\rightarrow\mathbf{p}+\frac{\mathbf{q}}{2}$ for the first and $\mathbf{p}\rightarrow\mathbf{p}-\frac{\mathbf{q}}{2}$ for the second term, yields the
desired result 
 \begin{eqnarray}\label{E3}
& &\int^{+\infty}_{-\infty}\frac{d\omega}{\pi}\big(-\frac{\textrm{Im}\chi_{JJ}^L(\omega,\mathbf{q})}{\omega}\big)\nonumber\\
&=&\sum_{\mathbf{p}}\frac{\mathbf{q}\cdot\mathbf{q}}{mq^2}\Big[1-\frac{\xi_{\mathbf{p}}}{E_{\mathbf{p}}}\big(1-2f(E_{\mathbf{p}})\big)\Big]\nonumber\\
&=&\frac{1}{m}\sum_{\mathbf{p}}\Big[1-\frac{\xi_{\mathbf{p}}}{E_{\mathbf{p}}}\big(1-2f(E_{\mathbf{p}})\big)\Big]\nonumber\\
&=&\frac{n}{m}.
 \end{eqnarray}

\subsection{Explicit Calculation of Diamagnetic Susceptibility}

The electromagnetic response kernel has been constructed so that there is no Meissner effect above $T_c$, only the $q_y^2$ coefficient in the series expansion of Eq.\ref{eq:P_Dia} is needed to calculate $\chi^{\textrm{dia}}$. The calculation is lengthy but straightforward, with the result that $\chi^{\textrm{dia}}$ can be written as the sum of two terms, 
$\chi^{\textrm{dia}}=\chi^{\textrm{dia}}_0+\delta\chi^{\textrm{dia}}$ 
 where
\begin{widetext}
\begin{eqnarray}
\chi^{dia}_0&=&-\displaystyle{\sum_{\bf k}}\Big(\frac{\partial\xi_{\bf k}}{\partial k_x}\Big)^2\frac{1}{12E_{\bf k}^7}\!\Bigg(3\Delta_{pg, \bf k}^2\Big(3m_{\bf k}^{-1}\xi_{\bf k}E_{\bf k}^2+v_{\bf k}^2\Big[\Delta_{pg,\bf k}^2\!-\!4\xi_{\bf k}^2\Big]\Big)\Big(1\!-\!2f(E_{\bf k})\Big)
\nonumber \\&&+6\Delta_{pg,\bf k}^2E_{\bf k}\Big(3m_{\bf k}^{-1}\xi_{\bf k}^3+3m_{\bf k}^{-1}\xi_{\bf k}\Delta_{pg,\bf k}^2-4\xi_{\bf k}^2v_{\bf k}^2+\Delta_{pg,\bf k}^2v_{\bf k}^2\Big)f^{\prime}(E_{\bf k})
\nonumber \\&&+6\xi_{\bf k}^2E_{\bf k}^2\Big(m^{-1}_{\bf k}\xi_{\bf k}E_{\bf k}^2+2\Delta_{pg, \bf k}^2v_{\bf k}^2\Big)f^{\prime\prime}(E_{\bf k})+2\xi^4_{\bf k}E^3_{\bf k}v^2_{\bf k}f^{(3)}(E_{\bf k})\Bigg)
~~~\rm{and}
\label{eq:chi1_Dia}
\end{eqnarray}
\end{widetext}
\begin{widetext}
\begin{eqnarray}
\delta\chi^{\textrm{dia}}&=&-\displaystyle{\sum_{\bf k}}(\frac{\partial\xi_{\bf k}}{\partial k_x})^2\!\frac{1}{12E_{\bf k}^7}\Bigg((\frac{\partial\phi_{\bf k}}{\partial k_y})^2(15\xi_{\bf k}^2\Delta_{pg,\bf k}^2[1-2f(E_{\bf k})]+30\xi_{\bf k}^2\Delta^2_{pg,\bf k}E_{\bf k}f^{\prime}(E_{\bf k})
\nonumber \\&&+6E_{\bf k}^2\Big(\xi_{\bf k}^4+\Delta_{\bf k}^4)f^{\prime\prime}(E_{\bf k})+2E_{\bf k}^3\xi_{\bf k}^2\Delta_{pg,\bf k}^2f^{(3)}(E_{\bf k}))
\nonumber \\&&-\frac{1}{6E_{\bf k}^7}\xi_{\bf k}\Delta_{pg,\bf k}v_{\bf k}\frac{\partial\varphi_{\bf k}}{\partial k_y}([6\xi^2_{\bf k}-9\Delta_{pg,\bf k}^2](1-2f(E_{\bf k}))+6(2\xi_{\bf k}^2-3\Delta_{pg,\bf k}^2)E_{\bf k}f^{\prime}(E_{\bf k})
\nonumber \\&&-6E^2_{\bf k}(\xi^2_{\bf k}-\Delta_{pg, \bf k}^2)f^{\prime\prime}(E_{\bf k})+2E^3_{\bf k}\xi^2_{\bf k}f^{(3)}(E_{\bf k}))
\nonumber \\&&-\frac{\Delta_{pg,\bf k}}{4E_{\bf k}^5}\frac{\partial^2\phi_{\bf k}}{\partial k_y^2}((\Delta^2_{pg,\bf k}-2\xi_{\bf k}^2)(1-2f(E_{\bf k}))+2(\Delta_{pg,\bf k}^2-2\xi^2_{\bf k})E_{\bf k}f^{\prime}(E_{\bf k})
\nonumber \\&&+2\xi_{\bf k}^2E_{\bf k}^2f^{\prime\prime}(E_{\bf k}))\Bigg)\label{eq:chi2_Dia}
\end{eqnarray}
\end{widetext}
Here we have separated terms so that the first
is the contribution independent of derivatives of the
gap form factor, while the second term introduces these
derivative contributions, which are notably absent in the
$s$-wave case. In the above equations we
have $v_{\mathbf{k}}^2 = 
\frac{1}{2} \big( 1 - \frac{\xi_{\mathbf{k}}}{E_{\mathbf{k}}} \big)$
and
$m_{\bf k}^{-1}\equiv\partial^2\xi_{\bf k}/\partial k_x^2$.

The expressions Eq.\ref{eq:chi1_Dia}-\ref{eq:chi2_Dia} are 
quite complicated, but general inferences can still be made on the 
expected behavior of the orbital susceptibility. 
We find that to a good approximation
Eq.\ref{eq:chi1_Dia} can be written as
\begin{eqnarray}
\chi^{\textrm{dia}}_0 &\approx&-\displaystyle{\sum_{\bf k}}\Big(\frac{\partial\xi_{\bf k}}{\partial k_x}\Big)^2\frac{1}{4\Delta_{pg,\bf k}^4}\Bigg(|\Delta_{pg,\bf k}|v_{\bf k}^2(1-2f(|\Delta_{pg,\bf k}|)) \nonumber \\
&+&2\Delta_{pg,\bf k}^2v_{\bf k}^2f^{\prime}(|\Delta_{pg,\bf k}|)\Bigg)
\label{eq:chi3_Dia}
\end{eqnarray}
and Eq.\ref{eq:chi2_Dia} is well approximated as
\begin{eqnarray}
\label{eq:chi4_Dia}\delta\chi^{\textrm{dia}})_0 \approx -\displaystyle{\sum_{\bf k}}\Big(\frac{\partial\xi_{\bf k}}{\partial k_x}\Big)^2\Bigg(\frac{1}{2}\Big(\frac{\partial\phi_{\bf k}}{\partial k_y}\Big)^2\frac{f^{\prime\prime}(|\Delta_{pg,\bf k}|)}{2|\Delta_{pg, \bf k}|} \nonumber \\
+\frac{1}{4\Delta_{pg, \bf k}^3}\frac{\partial^2\varphi_{\bf k}}{\partial k_y^2}\Big(|\Delta_{pg,\bf k}|(1-2f(|\Delta_{pg,\bf k}|)) \nonumber \\
+2\Delta_{pg,\bf k}^2f^{\prime}(|\Delta_{pg,\bf k}|)\Big)\Bigg)
\end{eqnarray}
It is seen from a numerical analysis that the largest contributing term to 
this last equation 
is the $(\partial\phi_{\bf k}/\partial k_y)^2$ piece.

\section{Numerical Results}
The sum of Equations (\ref{eq:chi3_Dia}) and
(\ref{eq:chi4_Dia}) provides a reasonable approximation to
the total orbital susceptibility
Eq.~(\ref{eq:chi1_Dia})-(\ref{eq:chi2_Dia}). Moreover 
all that is needed is a the temperature dependent
pseudogap parameter (and the chemical potential).
It is our intention to first present results that are 
independent of microscopic details so the plots are
representative of general theories in which we treat
$\Delta_{pg}$ and $\mu$ as variables.
We will simultaneously consider $s$ and $d$-wave
pairing in both jellium and tight
binding lattice models. For the latter we consider
the simplest dispersion

\begin{eqnarray}
\xi_{\bf k}=-2t(\cos(k_x)+\cos(k_y))-4t^{\prime}\cos(k_x)\cos(k_y)-\mu\label{eq:lattice_Dia}
\end{eqnarray}
along with more realistic fits to the cuprate bandstructure via ARPES data. 

\begin{figure}
\begin{center}
\includegraphics[width=3.25in,clip=true]
{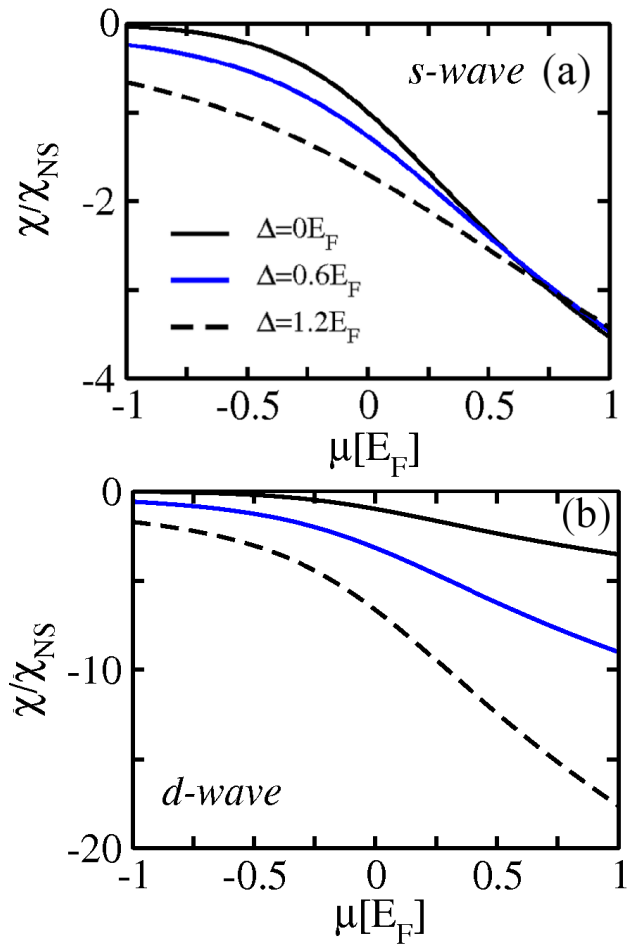}
\caption[] {The orbital susceptibility for a jellium dispersion
with an s-wave (upper) and d-wave gap (lower).
We consider
$\Delta_{pg}=0$,$0.6$, and $1.2E_F$. The curves are normalized by
$\chi_{NS}$, which is defined as the magnitude of the normal state
value of $\chi^{\textrm{dia}}$ at $\mu=0$. There is a finite
orbital susceptibility even in the normal state. The diamagnetism
is enhanced for all $\mu$ as size of the gap is increased. 
The $d$-wave gap symmetry results in a larger
$\chi^{\textrm{dia}}$ than in the $s$-wave case.}
\label{fig:chiSJellium_Dia}
\end{center}
\end{figure}

\subsection{Jellium models with $s$ and $d$-wave pairing}

We begin by studying a jellium dispersion and an $s$-wave gap 
which avoids the complications of lattice effects and gap nodes. In Fig.\ref{fig:chiSJellium_Dia}(a) we plot
the orbital susceptibility from Eq.\ref{eq:chi_Dia}. Curves are shown for $\Delta_{pg}=0,0.6,$ and $1.2E_F$ and normalized by $\chi_{NS}$, defined as the absolute value of the normal state orbital susceptibility when $\mu=0$. The system exhibits diamagnetism even in the normal state, reflecting the well-known Landau orbital susceptibility corresponding to
minus one third of the Pauli susceptibility.  The degree of diamagnetism is increased modestly as the gap size increases; this numerically confirms the intuition that the bosonic term in Eq.\ref{eq:P_Dia} results in an enhancement of $\chi^{\textrm{dia}}$.

\begin{figure}
\includegraphics[width=3.25in,clip]
{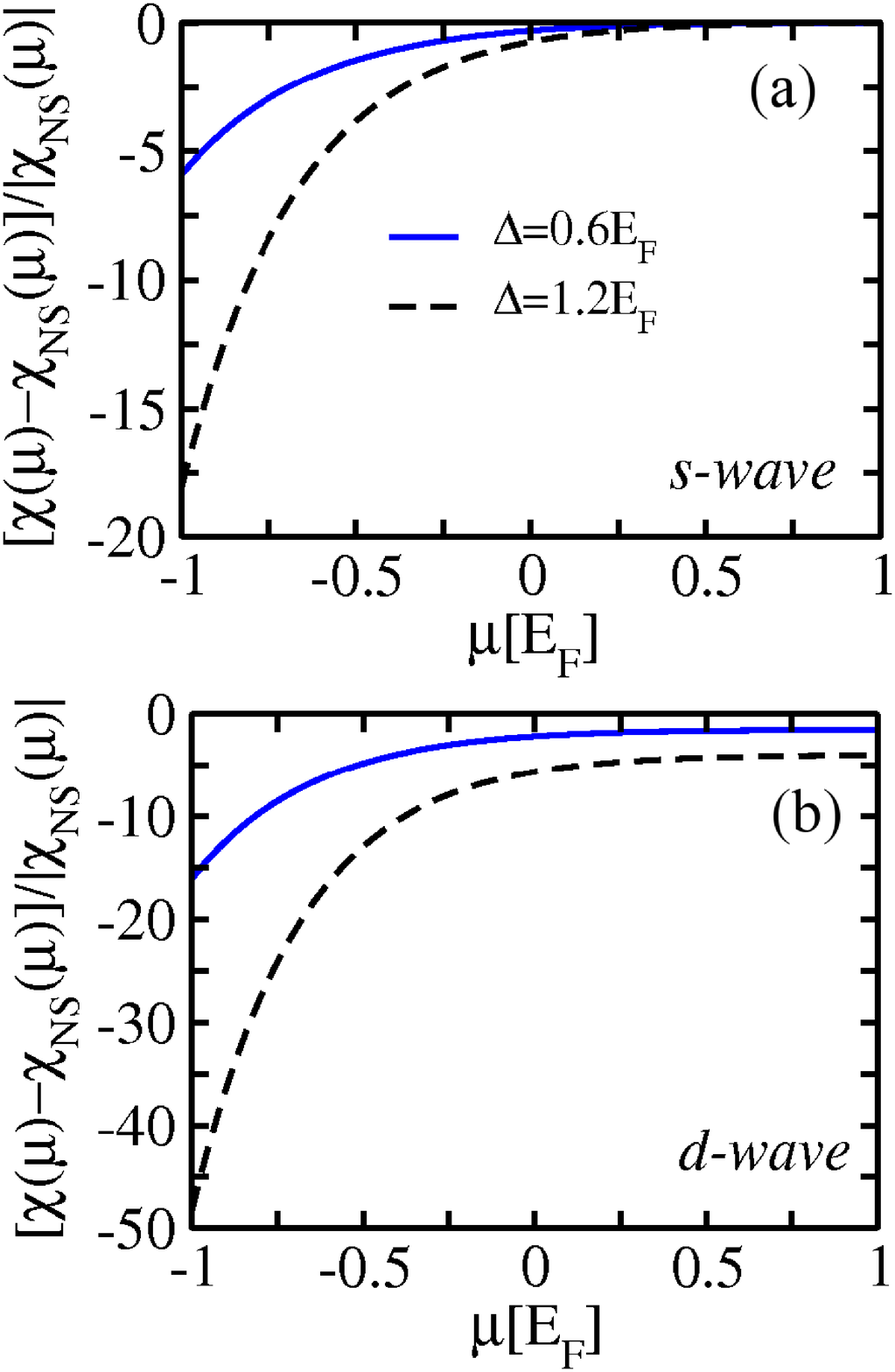}
\caption{Change in the orbital susceptibility with 
varying pseudogap for a jellium
dispersion. Plotted is the relative difference between the pseudogap and normal
state $\chi^{\textrm{dia}}$ as a function of chemical potential $\mu$ for
$\Delta=0.6,$ and $1.2E_F$. Upper panel is for
an $s$-wave and lower panel for a $d$-wave gap. The presence of
a pseudogap results in an increase of $\chi^{\textrm{dia}}$
 for all values of $\mu$.
The trends as a  function of $\mu$ and $\Delta$ are the same but the
overall magnitude of the enhancement is more significant for the $d$-wave
case.}
\label{fig:chidiffJ_Dia}
\end{figure}

For the case of a $d$-wave gap with
a jellium dispersion we use $\Delta_{pg,\bf k}=\Delta_{pg}\cos2\phi$ where $\phi$ is the azimuthal angle. The resulting plot
is shown in Fig.\ref{fig:chiSJellium_Dia}(b).
The curves are again displayed for $\Delta_{pg}=0,0.6,$ and $1.2E_F$ and normalized by $\chi_{NS}$. In contrast to the $s$-wave case, where $\chi$ grows as large as $\chi^{\textrm{dia}}\propto 3.5\chi_{NS}$, $\chi^{\textrm{dia}}\propto 17.5\chi_{NS}$ in the $d$-wave case, is roughly an order of magnitude larger than the normal state orbital susceptibility.

 We understand this contrast between the $s$ and $d$-wave order parameters as physically resulting from the increased pair size of the latter 
\cite{Benfatto_Dia,ChenPhD_Dia},
as reflected in the generic expression for orbital diamagnetism 
(Eq.\ref{eq:chiGL_Dia}). The increased pair size reflects the nodal structure of the gaps; the correlation length of the pairs (which is \textit{not} necessarily the same order of magnitude as the pair size) has a similar increase due to the nodes. It can
be argued that the correlation length $\xi\propto\Delta_{FS}^{-1}$ where $\Delta_{FS}$ is the average of the gap along the Fermi surface. The quantity $\Delta_{FS}$ will always be less relative to an $s$-wave order parameter as long as the gap maximum is set to the same value as the $s$-wave gap that it is being compared against.

Fig.\ref{fig:chidiffJ_Dia} shows
the change in the orbital susceptibility associated with
a pseudogap for the case of jellium. Here we plot
$\delta\chi_{NS}=
\chi(\mu)-\chi_{NS}(\mu)$ normalized by $|\chi_{NS}(\mu)|$ for the gap values $\Delta_{pg}=0.6,$ and $1.2E_F$, where the normal state
susceptibility is
at $\mu =0$. We consider 
$\Delta_{pg}=0.6$ and $1.2E_F$ for the blue and black-dashed curves respectively. 
Fig.\ref{fig:chidiffJ_Dia}(a)
shows $\delta\chi_{NS}$ for an $s$-wave order parameter
and
Fig.\ref{fig:chidiffJ_Dia}(b)
plots the counterpart figure,
$\delta\chi_{NS}$, for the $d$-wave case. For both the
$s$ and $d$ wave cases when the filling is decreased (so the
the normal state Landau orbital susceptibility is smallest)
then the relative effect of the pseudogap appears more
prominently.
As anticipated, the overall diamagnetic contribution due to the $d$-wave order parameter is significantly larger than the contribution occuring for an $s$-wave pseudogap.

In summary, for a jellium dispersion, we have seen that (i) the presence of a pseudogap leads to an enhancement of the normal state diamagnetism for most values of the chemical potential $\mu$. Critically, (ii) a $d$-wave order parameter results in a significantly larger diamagnetism than in the case of an $s$-wave order parameter.

\subsection{Tight-binding lattice}

\begin{figure}
\begin{center}
\includegraphics[width=3.7in,clip=true]{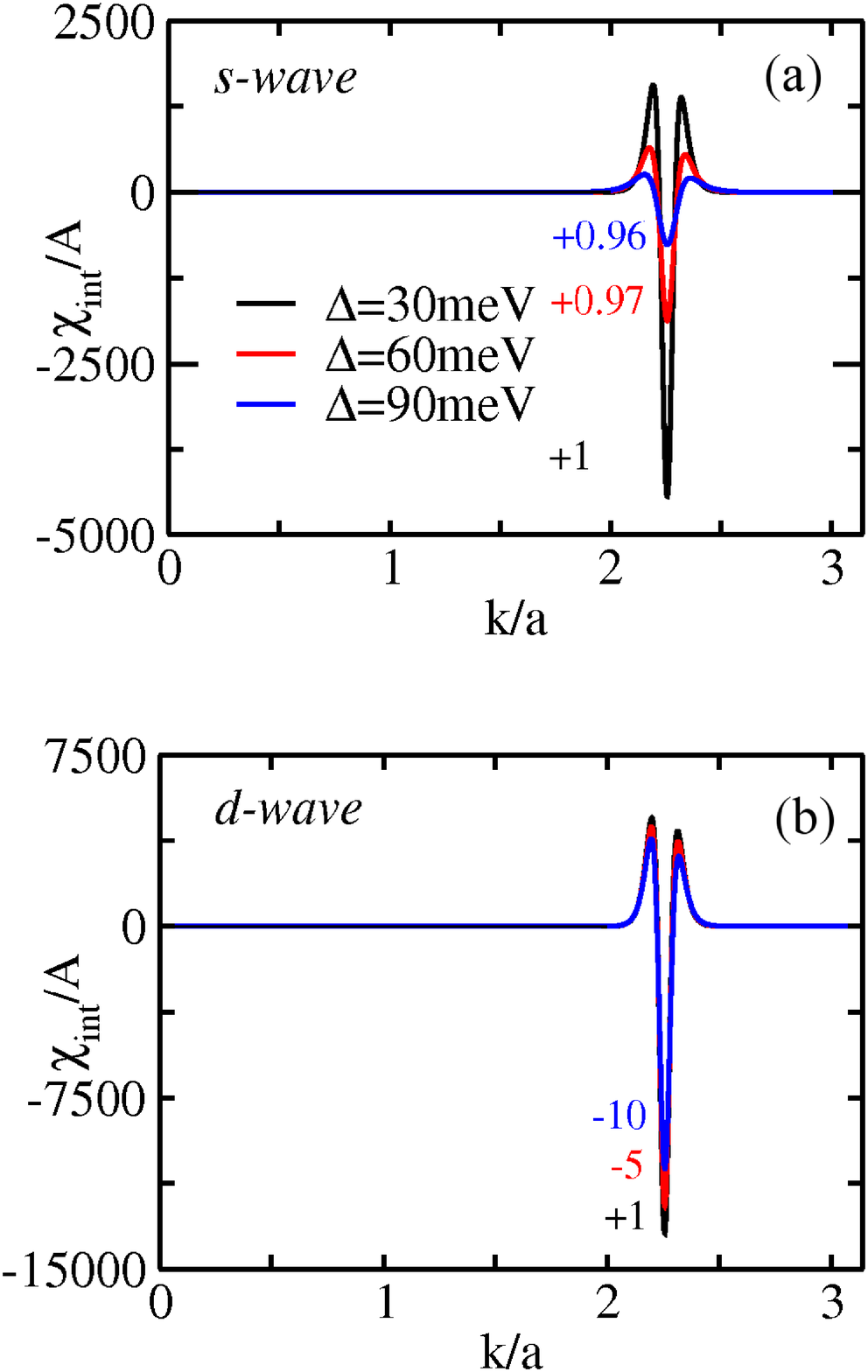}
\end{center}
\caption[]
{The orbital susceptibility integrand for a tight-binding
dispersion at an angle $7\pi/36$ in the $k_x$-$k_y$ plane. (top) The integrand 
for 
an s-wave order parameter of value $\Delta_{pg}=30,60,$ and $90$meV. The
curves are normalized by $A$, defined such that the area under the $\Delta_
{pg}=30$ meV curve is unity. The curve for all values of $\Delta_{pg}$ results
 in a paramagnetic contribution. The total area normalized by $A$ of each 
curve
is written alongside of it; the area takes the values $1$, $0.97$, and $
0.96$ for $\Delta_{pg}=30,60$,$90$meV respectively. (bottom) The same
plots for
a $d$-wave order parameter.
The curve for $\Delta_{pg}=60$ and $90$ meV results
in an overall diamagnetic contribution. The total area normalized by $A
$ of each curve is written alongside of it; the area takes the values $1$, $
-5$, and $-10$ for $\Delta_{pg}=30,60$,$90$meV respectively.}
\label{fig:integrand_Dia}
\end{figure}

It is useful to understand the integrand before addressing the
entire expression for the orbital susceptibility in
a tight binding lattice.
Figure \ref{fig:integrand_Dia} plots the integrand associated with
Eq.\ref{eq:chi1_Dia}-\ref{eq:chi2_Dia}, as a function of radial distance in the $k_x$-$k_y$ plane (normalized by the lattice spacing $a$). We consider three different values of the gap ($\Delta_{pg}=30,60$, and $90meV$). Each of the curves is normalized by a number $A$ selected so that the curve corresponding to the smallest gap $\Delta_{pg}=30$meV has area unity; the area of the curves is noted beside each.  All curves correspond to a cut at $35$ degrees above the $k_x$ axis. Fig.\ref{fig:integrand_Dia}(a) shows the integrand corresponding to an $s$-wave gap
and
Fig.\ref{fig:integrand_Dia}(b)
corresponds to the $d$-wave case.

For the former these cuts result in paramagnetic contributions to $\chi^{\textrm{dia}}$ for all values of $\Delta_{pg}$ and the extent of the contribution is largely unchanged as $\Delta_{pg}$ increases: despite a tripling of the gap size, the displayed
areas change by no more than $4\%$ and lead to a paramagnetic
contribution.

The corresponding curves for a $d$-wave order parameter are displayed in 
Fig. \ref{fig:integrand_Dia}(b). Although the
form of the curve does not change qualitatively as the gap size is changed, the initial area of the curve is sufficiently small such that minor changes to the cuts result in significant changes to the nature of the overall contribution to $\chi^{\textrm{dia}}$: as the gap increases from $30$ to $90$meV, the overall contribution changes from being paramagnetic to being $10$ times as large and diamagnetic. This trend is driven by the $d$-wave gap symmetry.

We now return to
Fig.\ref{fig:chi_NLattice_Dia} which plots the density of states
versus energy (a) and the orbital susceptibilities in the
gapless normal state (b) and in the presence (c) of an $s$- or(d)
a $d$ wave pairing. The last three panels are plots
as a function of band filling or chemical potential and
the units are in terms of
$t$, the nearest neighbor coupling which is taken to be $300~meV$.
Panel (b) shows that when 
$|\mu|$ is large, the density of holes or electrons is sufficiently small such that the system behaves similarly to the jellium case and $\chi^{\textrm{dia}}$ leads to the usual Landau orbital susceptibility result. There is a competition
\cite{Vignale_Dia}
, however, between the $E_F$ and $t$ energy scales, so that once $|E_F|/t\leq 1.8$ the system becomes entirely paramagnetic. The term that contributes to this paramagnetism is weighted by the density of states evaluated at the Fermi surface and so it is dramatically enhanced by the Van-Hove point at $E_F=0$. 
It is clear from the figure that the pseudogap-enhanced paramagnetism is
substantial but only for the $d$-wave case.

\begin{figure}
\begin{center}
\includegraphics[width=3.25in,clip=true]{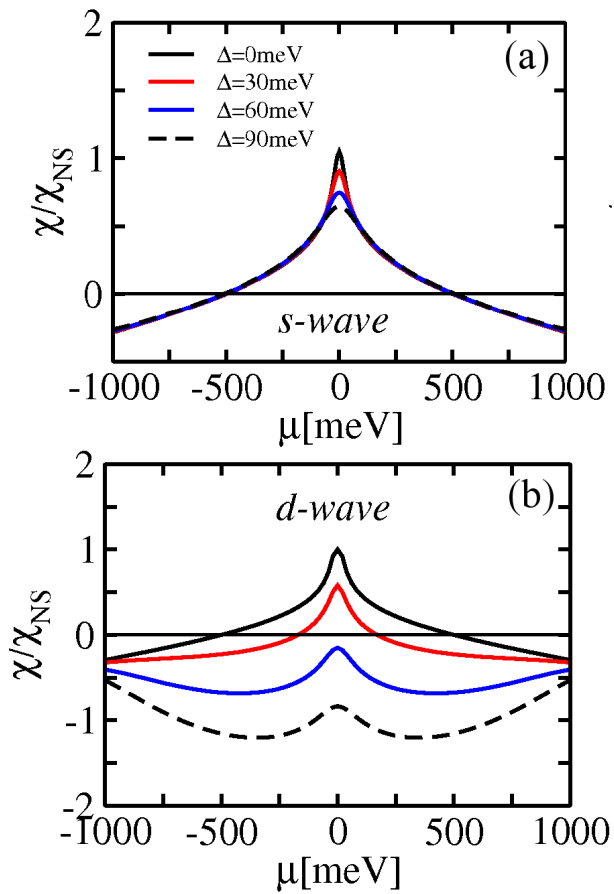}
\caption[]
{(left) The orbital susceptibility for a tight-binding dispersion with an s-wave order parameter. The normalization $\chi_{NS}$ is defined as the value of $\chi^{\textrm{dia}}$ in the normal state when $\mu=0$. The curves correspond to gap sizes $\Delta_{pg}=0,30,60,90$ meV. There is very little change of $\chi^{\textrm{dia}}$ regardless of the gap size. (right) The 
orbital susceptibility for a tight-binding dispersion with a d-wave order parameter normalized by $\chi_{NS}$. The $d$-wave form factor and the resulting enhancement in the size of the pairs leads to a significant enhancement of the diamagnetism for ranges of $\mu$ away from the Van-Hove point as the gap size increases.}\label{fig:chiLattice_Dia}
\end{center}
\end{figure}

Fig.\ref{fig:chiLattice_Dia}(a) 
presents a plot of the orbital susceptibility for an $s$-wave gap and a tight-binding dispersion normalized by $\chi_{NS}$, which is defined 
as in previous figures. The four curves correspond to the gap values $\Delta_{pg}=0,30,60,$ and $90meV$. The lattice dispersion Eq.\ref{eq:lattice_Dia} is used with $t=-300meV$ and $t^{\prime}=0$. The $s$-wave
pairing gap has little effect on $\chi^{\textrm{dia}}$, with all four curves tending to overlap except near the Van-Hove point at $\mu\approx0$. The curves are entirely paramagnetic for a significant range of 
filling and the diamagnetic features at other filling are relatively small: an $s$-wave gap on a lattice is not enough to capture the observed experimental effects. 

The corresponding figure for a $d$-wave gap is displayed in Fig.\ref{fig:chiLattice_Dia}(b). Gap and dispersion parameters that are identical to the $s$-wave parameters are used here. It is immediately seen that there is a significant enhancement of the diamgnetic susceptibility as the magnitude of the $d$-wave order parameter is increased. The Van-Hove point suppresses the extent of the diamagnetism and results in paramagnetism for small gap values and fillings near $\mu=0$; nevertheless, the diamagnetism persists over a wide range of filling for each of the gap values considered here. For fillings that are near mid-band 
(i.e. $\mu\approx 500meV$), it is seen that there is a substantial increase in $\chi^{\textrm{dia}}$ as compared to the normal state once a finite $\Delta_{pg}$ is considered. The extent of this enhancement increases as one moves away from the Van-Hove point, mirroring the experimentally observed trend of the anomalous diamagnetism having the greatest impact in underdoped samples.

\begin{figure}
\begin{center}
\includegraphics[width=3.25in,clip=true]{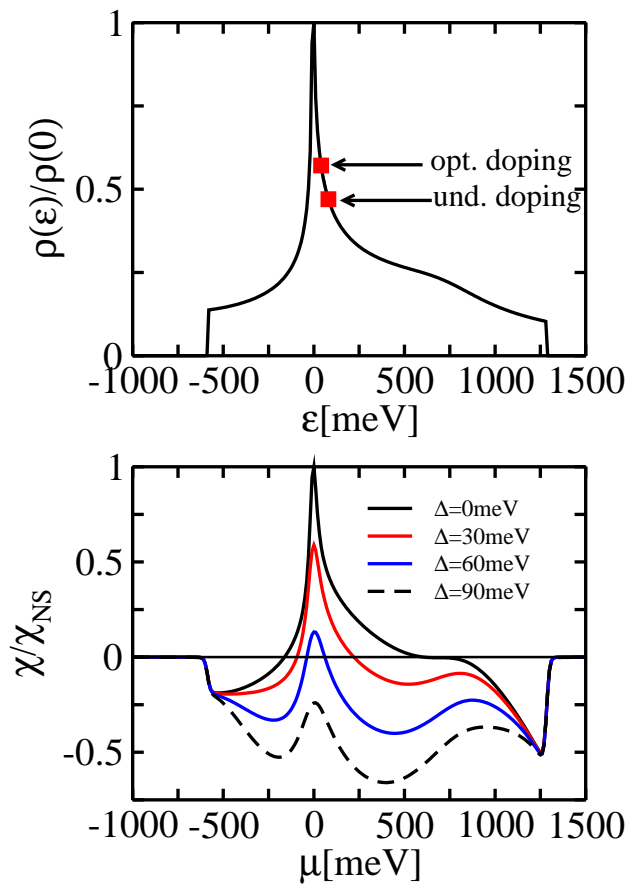}
\caption[]
{The behavor of the diamagnetism in a cuprate-like bandstructure. (upper panel) 
The model density of states showing that
the Van-Hove point occurs at negative energies. Optimally doped samples have chemical potentials closer to the Van-Hove point than underdoped samples, as indicated by the arrows. (lower panel) 
The diamagnetic ssuceptibility as a function of $\mu$ for $\Delta_{pg}=0,30,60,$ and $90meV$ normalized by the value of the normal state $\chi^{\textrm{dia}}$ at $\mu=0$. }
\label{fig:doping_Dia}
\end{center}
\end{figure}

\begin{figure}
\includegraphics[width=2.8in,clip]
{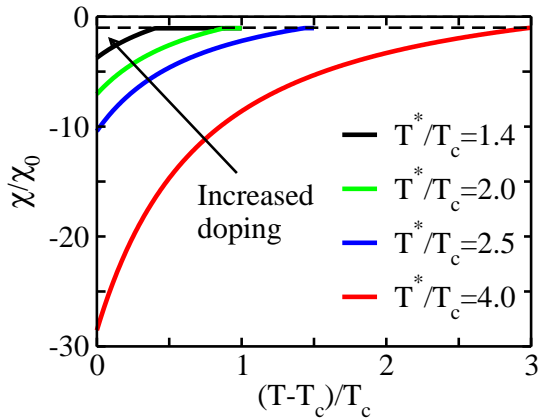}
\caption{
The diamagnetism as a function of temperature for different ``hole
concentrations" parameterized via $T^*/T_c$. The dashed line shows the normal state value of t
he susceptibility.
This diamagnetism is to be associated with the large transport
response from pairs or bosons near condensation. We contrast the
present
precursor superconductivity scenario with the more
more widely discussed phase fluctuation
approach where the diamagnetism is associated with mesoscopic
supercurrents.}
\label{fig:3}
\end{figure}


We turn now to more realistic parameters for the bandstructure of
the cuprates in 
Fig.\ref{fig:doping_Dia}.
Here we use the
ARPES-derived parameters \cite{Fischer_Dia} for Bi2212. 
With next nearest neighbor contributions one sees                                  the broad trends are consistent with the earlier results. Note that the finite $t^{
\prime}$ breaks particle hole symmetry and shifts the Van-Hove point in the density of states.

Fig.\ref{fig:doping_Dia}(a) indicates a typical density of states for a cuprate system, along with the estimated positions for the
chemical potentials. 
Here the Van-Hove point is located at negative energies. The general trend is that optimally doped systems will be closer to the Van-Hove point than underdoped systems (\cite{Fischer_Dia}).
Interestingly, this along with the stronger pseudogap in the
underdoped case leads us to anticipate
that underdoping enhances diamagnetism.
The resulting orbital susceptibility is plotted in Fig.\ref{fig:doping_Dia}(b)
as a function of band filling  for $\Delta=0,30,60,$ and $90$meV. The curves are qualitatively similar to previous figures with the normal state $\chi^{\textrm{dia}}$ having both regions of paramagnetism and diamagnetism and becoming entirely diamagnetic for a sufficiently large value of $\Delta_{pg}$.

\subsection{Temperature Dependence of Diamagnetism}

The above discussion has been quite general; we have essentially
explored the consequences of Eq.~(\ref{eq:chi_Dia}) in all its generality.
One can then inquire as to how this applies to the high temperature
superconductors. 
This requires that we
establish the parameters
$\Delta_{pg}(T)$  and the fermionic chemical potential, $\mu(T)$.
Here, for definiteness we use our
preformed pair scenario \cite{ourreview} which is based on
stronger than BCS attractive interactions (consistent with small pair size and anomalously high pairing onset temperature $T^*$).
Once the pseudogap and chemical potential parameters are self consistently obtained
\cite{ourreview}, one
accomodates a variety of dopings, by effectively fitting \cite{ourreview} the attractive
interaction to match $T^*$ and $T_c$. For definiteness, we
presume the band dispersion is associated with
$t=300 meV$
and $\mu
=-1.75t$, chosen somewhat away from the Van Hove point.

We present temperature dependent plots of the diamagnetic response 
$\chi^{\textrm{dia}}$,
in
Figure~\ref{fig:3} for 
four
different dopings. Each curve is normalized by its normal state value, which is separately plotted as a dashed line. Independently of the particular parameters that are used, it is seen
that the magnitude of
$\chi^{\textrm{dia}}$ is enhanced even at temperatures well above $T_c$.
Importantly,
this diamagnetism is
not restricted to two dimensional models, as in fluctuation theories.
Experiments as well re-enforce
three dimensional
critical behavior \cite{Kamal,Overend,Pureur}.

In this way the present physical picture contrasts with the traditional
fluctuation approach, in which one might expect large diamagnetism
but only in the narrow critical regime.
Here, it is the stronger than BCS
attraction which stabilizes these pair degrees of freedom
(up to high temperatures $ T \approx T^*$)
rather than the low dimensionality
\cite{LarkinVarlamov}.
Importantly, in the present theory we compute the total
conductivity based on the pseudogap self
energy, not the fluctuation corrections and in this way are 
analytically able to establish compatibility with the conductivity
sum rules.

\begin{figure*} \includegraphics[width=6.5in,clip]
{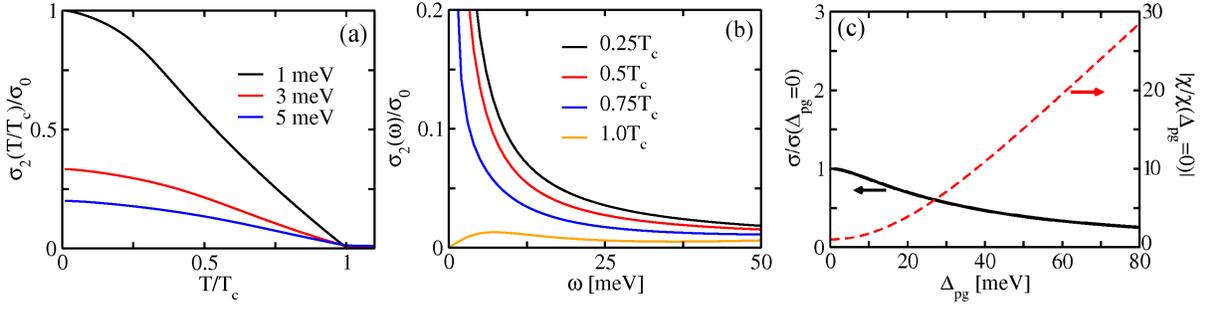}
\caption{
Numerically generated plots of
(a) the imaginary conductivity $\sigma_2$ as a function of temperature at constant energy $\omega$. The curves are normalized by $\sigma_0$, the value of $\sigma_2$ at zero temperature with $\omega=1meV$. (b) The imaginary conductivity $\sigma_2$ as a function of frequency for constant temperature $T$. The curves are again normalized by $\sigma_0$.
(c) The conductivity and diamagnetic
susceptibility as functions of $\Delta_{pg}$ at $k_BT=15meV$. } \label{fig:2}
\end{figure*}

\section{Comparison between Diamagnetism and
Conductivity}

The orbital susceptibility is not the only transport property which
can be associated with the pseudogap self energy of
Eq.~(\ref{eq:1}). We have previously discussed the optical
and THz conductivity \cite{OurAC,LongConduct}. 
Here we concentrate on the challenges raised by
recent experimental papers \cite{Bilbro,Bilbro2} which have pointed out
the seemingly contradictory
behavior implicit in the dissipative conductivity and
the orbital susceptibility.  
The authors of Ref.~\onlinecite{Bilbro}
deduce ``However, if the diamagnetism signal above
$T_c$ is solely due to superconducting correlations then it is a
well posed challenge to explain the lack of straightforward
correspondence to conductivity."
In this section, we address this challenge by looking simultaneously
at the linear diamagnetic response
and the dissipative conductivity.

In this regard, it is important here to incorporate
the ordered phase. We
can anticipate that
we now have two distinct contributions to the fermionic
self energy $\Sigma({\bf k}, \omega)$
\begin{eqnarray}\label{eq:16}
\Sigma({\bf k},\omega) &=& \Sigma_{pg,{\bf k}} + \Sigma_{sc,{\bf k}} \nonumber \\
&=& -i \gamma + \frac{\Delta_{pg,\bf k}^2}{\omega+\xi_{\textbf{k}}+i\gamma} +
\frac{\Delta_{sc,\bf k}^2 }{\omega+\xi_{\textbf{k}}} \ .
\end{eqnarray}
The first of these is associated with the normal state and the second
with the condensate.

Following the analysis of this paper (above $T_c$) we may similarly
arrive at the below $T_c$ counterpart directly from this self energy. Here for
simplicity we rewrite 
$\tensor{P}(Q)$
in the regime of very weak dissipation ($\gamma \approx 0$)
and $s$-wave pairing where the
behavior is more physically transparent. 
\begin{eqnarray}
& &\tensor{P}(\omega,\mathbf{q})=\sum_{\mathbf{k}}\frac{\mathbf{k}\mathbf{k}}{m^2}\Big[\frac{E_++E_-}{E_+E_-}\big(1-f_+-f_-\big)\nonumber\\                                                                   & &\times\frac{E_+E_--\xi_+\xi_--\delta\Delta^2}{\omega^2-(E_++E_-)^2}-\frac{E_+-E_-}{E_+E_-}\nonumber\\
& &\times\frac{E_+E_-+\xi_+\xi_-+\delta\Delta^2}{\omega^2-(E_+-E_-)^2}\big(f_+-f_-\big)\Big],
\label{eq:5}
\end{eqnarray}
where $f_{\pm}=f(E_{\pm})$ and $\delta\Delta^2=\Delta^2_{\textrm{sc}}-\Delta^2_{\textrm{pg}}$,
$\xi_{\pm}=\xi_{\textbf{k}\pm\textbf{q}/2}$,
and
$E_{\pm}=E_{\mathbf{k}\pm\mathbf{q}/2}$.
The transverse sum rule can be shown to be
precisely satisified, below as well as above $T_c$. The longitudinal
sum rule cannot be easily proved below $T_c$, (due to
collective mode effects) although it is analytically
satisied in the normal state as shown earlier in this paper.

Our interest here is on
contrasting the diamagnetism with the dissipative conductivity
 \begin{eqnarray}
\sigma_2(\omega)=\displaystyle{\lim_{\textbf{q}\rightarrow0}}\textrm{Re}
\Bigg[\frac{
P_{xx}(\textbf{q},\omega)+(n/m)_{xx}}{i\omega}\Bigg]\label{eq:conddef}
\label{eq:1f}
\end{eqnarray}
where the imaginary counterpart corresponds to the real part of the
frequency dependent conductivity $\sigma_1$.
Note that Eqs.~(\ref{eq:chi_Dia}) and (\ref{eq:1f}) are superficially
rather similar, but importantly very different. 
This difference derives from the different denominators and in a
related fashion,
the order of $\textbf{q}, \omega \rightarrow 0$ limits.

We stress that
our below $T_c$ extension can be derived microscopically 
\cite{ourreview,Chen2,Kosztin2}.
However, the present more general approach based on the self energy indicates that
the results
should be rather generic.
In support of this below-$T_c$ generalization of the self energy, it
is straightforward to see from Eq.~(\ref{eq:16})
how the Fermi arcs
will collapse to point nodes as soon as the system passes
below $T_c$ \cite{FermiArcs} and that the general 2-gap
phenomena, in which the nodal region is sensitive to
$\Delta_{sc}$ while the antinode region exhibits little
temperature dependence \cite{ourarpes}
are all direct consequences. In the same way one can
explore the optical conductivity \cite{OurAC} and
THz conductivity \cite{LongConduct}, all of which are
highly constrained by the simple self energy
expression in Eq.~(\ref{eq:1}) and its extension
below $T_c$ in Eq.~(\ref{eq:16}).
In addition, these
expressions for the self energy lead to
results for the specific heat \cite{Chen4}
and serve to constrain the tunneling characteristics,
including quasi-particle interference \cite{ourqpi}.
Because the measured diamagnetism seems to be
well correlated with the Nernst effect, it is important to note that
a large Nernst response was addressed earlier and argued \cite{TanNernst}
to be associated with pre-formed pairs, as distinguished
from normal state vortices.

It is of interest then to compare the behavior of the conductivity
and the diamagnetic response.
Fig.\ref{fig:2} displays our  results
for the cuprate models, based on
Eqs.~(\ref{eq:chi_Dia}) and (\ref{eq:conddef}).
Fig.\ref{fig:2}(a) shows how the normal state $\sigma_2(\omega)$
behaves
as a function of temperature for three different low frequencies
and 
Fig.\ref{fig:2}(b) plots the imaginary conductivity $\sigma_2$ as a function of frequency for
a range of different temperatures.
At roughly $T_c$, we find
that $\sigma_2$ shows a sharp upturn at low $\omega$.
Below $T_c$ we find the expected superfluid 
frequency dependence $\sigma_2 \propto n_s/\omega$.

We find that $\sigma_2(\omega)$ is modestly increasing
with increasing $\omega$ above $T_c$
as seen in experiment, albeit slightly away from the transition.
Experimental studies do reveal a
small 10-15K range where a fluctuation contribution
is visible.
But except in this  narrow temperature region, the observed behavior
is not compatible with that expected of a phase fluctuation
contribution, where a $\omega^{-1}$ dependence would occur \cite{Hanke},
presumably over the same range of temperatures as the enhanced
diamagnetic response. Both derive
from the same mesocopic supercurrents. 
The absence of significant $\sigma_2$ contributions above $T_c$ in the present
theory is related to the fact that
the pseudogap does not
contribute to
the superfluid density. Our derivation of the current-current
correlation function centered on this important constraint.
Thus, it should not be surprising that we find
little signature of the
superfluid density in $\sigma_2$ of the normal phase.
Here, moreover, fluctuation effects are not present since
ours is basically a mean field approach.
As speculated in Ref.~\onlinecite{Bilbro}, one should
distinguish these near $T_c$ critical fluctuations from
preformed pairs that persist to much higher temperatures
closer to $T^*$.

It is notable that even $\sigma_1$ is suppressed in
the low frequency regime, when a pseudogap is present. This is because
there are fewer fermions available to contribute to
transport;  their number is reduced since they are tied up into
pairs.
However, once the frequency is sufficiently high
to break the pairs into individual fermions,
the conductivity rises, leading \cite{OurAC} to a second peak  
at higher $\omega \approx 2 \Delta$) in $\sigma_1(\omega)$.
One can think of this two component behavior (which is
observed experimentally \cite{Basov3}) as reflecting 
a transfer of spectral weight as implied by the conductivity f-sum rule.
The behavior of
$\sigma_2(\omega)$, shown in Fig. \ref{fig:2}(b) is rather similarly
constrained by these same effects.
On general principles, $\sigma_2$ must vanish at strictly zero frequency as long
as the system is normal. 
Here, too,  the low frequency behavior is suppressed by the
presence of a pseudogap as a result of the gap-induced decrease in the number of carriers.
The second peak in $\sigma_1$ 
which reflects the breaking of pairs, leads,
via a Kramers Kronig transform
to a slight depression in $\sigma_2(\omega) $ in this  
frequency regime.
Thus, $\sigma_2(\omega)$ is significally reduced
relative to a traditional Fermi liquid.

Fig.\ref{fig:2}(c)
compares the effects of a $d$-wave pseudogap on the diamagnetism 
and dc conductivity.
Here we presume that the chemical potential of
the underdoped cuprates is somewhat away from the Van Hove
points, to avoid the large normal state \textit{paramagnetism}
\cite{Vignale}.
The left and right hand axes plot the zero frequency conductivity as
a function of varying pseudogap energy scale $\Delta_{pg}$ and the orbital
susceptibility with varying $\Delta_{pg}$ respectively.
Once pre-formed pairs are present,
the diamagnetic contribution is significantly enhanced, relative to
the very weak band diamagnetism found at $\Delta_{pg}=0$.

These observations suggest an important
anti-correlation between the dc conductivity and the orbital susceptibility in the pseudogap phase,
which is shown in
Fig.\ref{fig:2}(c).
The dc conductivity is reduced with increasing $\Delta_{pg}$
due to the opening of a gap in the
fermionic spectrum. 
By contrast the existence of bosonic degrees of freedom (in the vicinity
of condensation) and $d$-wave pairing
allows for an enhanced diamagnetic response.

\section{Conclusion}

Because of the rather widespread adoption 
\cite{Malypapers,Normanphenom,SenthilLee,Chubukov2}
of a broadened BCS form for
the pseudogap self energy
(as given in Eq.~(\ref{eq:1})), it seems appropriate to explore
the consequences for transport, and in particular diamagnetism.
We have done so here 
in a fashion which is rather independent of the microscopic
details. 
%
The logic we followed is straightforward. We used the
absence of a normal state Meissner effect to constrain
the $q=0, \omega =0$ component of the current-current
correlation function $\overleftrightarrow{P}(0.0)$. We extended 
this correlation function to finite 4-vector arguments 
$\overleftrightarrow{P}(Q)$ in a fashion
which analytically satisfies the f-sum rules,
and yields known results for limiting cases. This
correlation function expanded for small wavevector, then,
directly yields the orbital susceptibility, associated with
the pseudogap self energy.

In this way, our numerical studies are expected to
yield rather generic results, a main
summary of which is presented in Figure 1. The orbital
susceptibility associated with a tightbinding
bandstructure is very different from more familiar Landau diamagnetism
\cite{Vignale_Dia}
and, importantly, near Van Hove singularities this
susceptibility is paramagnetic.
However, the effect of a pseudogap can lead to
large negative corrections to the bandstructure predictions,
for the $d$ wave case. That such diamagnetism is
present only for $d$-wave symmetry is due in part
to the presence of nodal fermions and to the extended
size of the $d$-wave pairs.

It is of interest to contrast our approach
with other theories in the literature.
Podolsky et al. \cite{Ashvin} and Eckl and Hanke \cite{Hanke} have
respectively applied  phase fluctuation approaches to address diamagnetism
and conductivity
experiments leading to several predictions:
(i) that
\cite{Hanke} the closely
related
$\sigma_2(\omega) \propto \omega^{-1}$, at sufficiently high $\omega$.
(ii) that
\cite{Ashvin},
the underdoped cuprates behave as a dilute vortex liquid over
a wide range of temperatures above $T_c$.
More recently, however, there has been some concern raised \cite{Bilbro2} about
this vortex
plasma model for the cuprates. A comparison of the conductance and
orbital susceptibility suggests that these
vortices must exhibit an anomalously large vortex
diffusion constant. We note that these concerns do not apply to
the present approach to transport.
Also notable is
a large body of work
on related vortex scenarios\cite{Leevortices,IoffeMillis},
as well as
theoretical studies \cite{Benfatto}
which address the superconducting fluctuation contribution to
conductivity and diamagnetism in the presence of current-current
interactions in the t-J model.

The calculations we present here 
are different primarily because we consider non-Fermi liquid
aspects to be dominant.
We have shown 
how a large diamagnetic (linear) response
derives in part from the well known
enhanced transport contribution associated with bosons near condensation.
At the same time, in the present theory, these bosons are meta-stable
at temperatures much higher than in the critical regime, away from where
standard fluctation theories apply. This high temperature
stability is due to
stronger-than-BCS attractive interactions.
We caution that this lattice diamagnetism is associated with two constraints:
that the pairing be $d$-wave and that the (near mid-band)
chemical potential
$\mu$ lie away
from the Van Hove singularities. When $\mu$ is at the
Van Hove point, the strong
paramagnetism of the normal state \cite{Vignale} inhibits
a diamagnetic response, as does $s$-wave pairing in a tight
binding band.

In summary, this preformed pair pseudogap scenario
leads to very different behavior for the conductivity and
the diamagnetism and this appears to bear on
recent, otherwise challenging
experiments \cite{Bilbro,Bilbro2}.  In the former case because
there are fewer fermions around to contribute
to the $\omega = 0$ conductivity (they are tied up into ``bosons")
this leads to
a reduction in the dissipative contribution to  the conductivity
Because we are working at effectively zero magnetic
field, we have not addressed diamagnetism associated with
a non-linear response, although this appears to be very
anomalous experimentally \cite{Ong2}.
We end by reiterating 
that our starting point, 
Eq.~(\ref{eq:1}), for deriving
$\sigma(\omega)$ and $\chi^{dia}$ is 
frequently adopted in the literature so our results
should have a wider applicability and generality.

\ \\

This work is supported by NSF-MRSEC Grant
0820054. We thank Hao Guo and Chih-Chun Chien for
valuable insights, and Dr. Guo for the analytic proof of
the longitudinal f-sum rule.

\ \\
\appendix

\section{Appendix A: Effects of finite $\gamma$ }

\begin{widetext}

The current-current correlation function will be discussed below in the normal state
in the presence of lifetime $\gamma^{-1}$, which is kept arbitrary in the analysis.

The starting point in the derivation of the current-current correlation function $\overleftrightarrow{P}(\textbf{q},i\Omega_m)$ is the phenomenological cuprate self-energy $\Sigma_{pg,K}$ as determined by ARPES experiments. It takes the form
\begin{eqnarray}
\Sigma_{pg,K}=-i\gamma\textrm{sgn}\omega_n+\frac{\Delta_{pg,\bf k}^2}{i\omega_n+\xi_{\bf k}+i\gamma\textrm{sgn}\omega_n}\label{eq:SE_THE}
\end{eqnarray}
where $K=(\textbf{k},i\omega_n)$ is a 4-vector, $i\omega_n$ is a fermionic Matsubara frequency, $\xi_{\bf k}$ is the normal state fermion dispersion, $\gamma^{-1}$ is the pseudogap pair lifetime, $\textrm{sgn}(..)$ denotes the sign function, and $\Delta_{pg,\bf k}\equiv\Delta_{pg}\varphi_{\bf k}$ where $\Delta_{pg}$ is the magnitude of the pseudogap and $\varphi_{\textbf{k}}$ is the gap form factor. Eq.\ref{eq:SE_THE} is the generalization of the phenomenological $\Sigma_{pg,K}$ to finite temperature, rather than the $i\omega_n\rightarrow\omega+i0^+$ limit that it is normally displayed in. The other physical constraint that will be used is the absence of the Meissner effect above $T_c$. This condition is expressed
\begin{eqnarray}
\overleftrightarrow{P}(0,0)+\frac{\overleftrightarrow{n}}{m}=0\label{eq:Meiss_THE}
\end{eqnarray}
and relates the zero momentum and frequency current-current correlation function to $\overleftrightarrow{n}/m$. Eq.\ref{eq:Meiss_THE} allows one to avoid issues of Ward identities and renormalization of the bare electromagnetic vertex when deriving $\overleftrightarrow{P}(0,0)$ because the tensor $\overleftrightarrow{n}/m$ is defined solely in terms of the normal state dispersion $\xi_{\bf k}$ and the Green's function $G_K$. By incorporating Eq.\ref{eq:SE_THE}, Eq.\ref{eq:Meiss_THE} can be used to calculate the electromagnetic vertex renormalization by first rewriting $\overleftrightarrow{n}/m$
\begin{eqnarray}
\frac{\overleftrightarrow{n}}{m}=2\displaystyle{\sum_{K}}\frac{\partial^2\xi_{\bf k}}{\partial\textbf{k}\partial\textbf{k}}G_K=-2\displaystyle{\sum_K}\frac{\partial\xi_{\bf k}}{\partial\bf k}\frac{\partial G_K}{\partial\bf k}\label{eq:NUM_THE}
\end{eqnarray}
where the second equality follows from an integration by parts. The derivative of the Green's function can be written in terms of a bare Green's function and Eq.\ref{eq:SE_THE}
\begin{eqnarray}
\frac{\partial G_K}{\partial\bf k}=-G_K^2\frac{\partial G_K^{-1}}{\partial\bf k}=G_K^2\Bigg(\frac{\partial\xi_{\bf k}}{\partial\bf k}+\frac{\partial\Sigma_{pg, K}}{\partial\bf k}\Bigg)\label{eq:GDER_THE}
\end{eqnarray}
Inserting Eq.\ref{eq:GDER_THE} into Eq.\ref{eq:NUM_THE} yields
\begin{eqnarray}
\frac{\overleftrightarrow{n}}{m}=-2\displaystyle{\sum_K}G_K^2\frac{\partial\xi_{\bf k}}{\partial\bf k}\frac{\partial\xi_{\bf k}}{\partial\bf k}\Bigg(1-\Delta_{pg,\bf k}^2{G^{\gamma}_{0,-K}}\Big)\label{eq:NUM2_THE}
\end{eqnarray}
where $G^{\gamma}_{0,K}$ is defined
\begin{eqnarray}
G^{\gamma}_{0,K}\equiv\frac{1}{i\omega_n-\xi_{\bf k}+i\gamma\textrm{sgn}\omega_n}
\end{eqnarray}
and takes the appearance of a normal state Green's function with a self-energy contribution due to a scattering process resulting in the $i\gamma$ factor. Eq.\ref{eq:NUM2_THE} implies that the zero momentum and zero frequency current-current correlation function is
\begin{eqnarray}
\overleftrightarrow{P}(0,0)=2\displaystyle{\sum_K}G_K^2\frac{\partial\xi_{\bf k}}{\partial\bf k}\frac{\partial\xi_{\bf k}}{\partial\bf k}\Bigg(1-\Delta_{pg,\bf k}^2{G^{\gamma}_{0,-K}}^2\Bigg)\label{eq:P_THE}
\end{eqnarray}
The key observation to make of Eq.\ref{eq:P_THE} is that (i) the phenomenological self-energy and (ii) ensuring the absence of a Meissner effect above $T_c$ were sufficient assumptions to calculate the renormalization of the bare electromagnetic vertex $\lambda_{\bf k}$ in the $\textbf{q},i\Omega_m=0$ limit, namely that the dressed electromagnetic vertex $\Lambda_{\bf k}$ is
\begin{eqnarray}
\lambda_{\bf k}=\frac{\partial\xi_{\bf k}}{\partial\bf k}\rightarrow \Lambda_{\bf k}=\frac{\partial\xi_{\bf k}}{\partial\bf k}\Bigg(1-\Delta_{pg,\bf k}^2{G^{\gamma}_{0,-K}}^2\Bigg)\label{eq:LAM_THE}
\end{eqnarray}
The transport properties derived from Eq.\ref{eq:LAM_THE} and its extension to finite momentum $\textbf{q}$ are of very general character in describing the pseudogap state. The remaining step is the extension of Eq.\ref{eq:P_THE} to finite $\textbf{q}$ and $i\Omega_m$ which will be accomplished at this stage by analogy to the BCS current-current correlation function but will be supplemented 
by an analytic proof of the 
tranverse f-sum rule that depends on the $i\Omega_m$ dependence of $\overleftrightarrow{P}(\textbf{q},i\Omega_m)$. One can utilize
the $\textbf{q}$ and $i\Omega_m$ dependences in BCS theory to
build further confidence by noting that the correction to the bare electromagnetic vertex is of the BCS-form but with opposite sign. Mirroring the $\textbf{q}$ and $i\Omega_m$ dependence of $\lambda_{\bf k}$, the Green's function, and the correction to $\lambda_{\bf k}$, the extension of Eq.\ref{eq:P_THE} is
\begin{eqnarray}
\overleftrightarrow{P}(\textbf{q},i\Omega_m)=2\displaystyle{\sum_K}G_KG_{K+Q}\frac{\partial\xi_{\textbf{k}-\textbf{q}/2}}{\partial\bf k}\frac{\partial\xi_{\textbf{k}-\textbf{q}/2}}{\partial\bf k}\Bigg(1-\Delta_{pg,\bf k}\Delta_{pg, \textbf{k}+\textbf{q}}G_{0,-K}^{\gamma}G^{\gamma}_{0,-K-Q}\Bigg)\label{eq:P2_THE}
\end{eqnarray}
The analogy between Eq.\ref{eq:P2_THE} and its BCS counterpart can be further exploited by introducing the function $F_{pg,K}$, defined as
\begin{eqnarray}
F_{pg,K}\equiv\Delta_{pg,\textbf{k}}G^{\gamma}_{0,-K}G_K\label{eq:FPG_THE}
\end{eqnarray}
While this is of similar form to the anomalous propagator $F_{sc,K}$, $F_{pg,K}$ 
\textit{does not} reflect phase coherent pairs. Inserting Eq.\ref{eq:FPG_THE} into Eq.\ref{eq:P2_THE} leads to a compact form 
\begin{eqnarray}
\overleftrightarrow{P}(\textbf{q},i\Omega_m)=2\displaystyle{\sum_K}\frac{\partial\xi_{\textbf{k}-\textbf{q}/2}}{\partial\bf k}\frac{\partial\xi_{\textbf{k}-\textbf{q}/2}}{\partial\bf k}\Bigg(G_{K}G_{K+Q}-F_{pg,K}F_{pg,K+Q}\Bigg)
\label{eq:32}
\end{eqnarray}
The proof of the transverse sum rule will be facilitated by working with the spectral representations of $G_K$ and $F_{pg,K}$, which are implicitly defined through the relations
\begin{eqnarray}
G_K&=&\displaystyle{\int}\frac{d\omega}{2\pi}\frac{A_G(\textbf{k},\omega)}{i\omega_n-\omega}\label{eq:SPEC_THE}\\
F_{pg,K}&=&\displaystyle{\int}\frac{d\omega}{2\pi}\frac{A_{F_{pg}}(\textbf{k},\omega)}{i\omega_n-\omega}
\end{eqnarray}
Inserting Eq.\ref{eq:SPEC_THE} into Eq.\ref{eq:P2_THE} leads to the result
\begin{eqnarray}
\overleftrightarrow{P}(\textbf{q},i\Omega_m)&=&2\displaystyle{\sum_{\textbf{k}}}\frac{\partial\xi_{\textbf{k}-\textbf{q}/2}}{\partial\bf k}\frac{\partial\xi_{\textbf{k}-\textbf{q}/2}}{\partial\bf k}\displaystyle{\int}\frac{d\omega d\omega^{\prime}}{(2\pi)^2}\Bigg(\frac{f(\omega)-f(\omega^{\prime})}{\omega-\omega^{\prime}+i\Omega_m}\Bigg)\\&&\times\Bigg(A_G(\textbf{k},\omega)A_G(\textbf{k}+\textbf{q},\omega^{\prime})-A_{F_{pg}}(\textbf{k},\omega)A_{F_{pg}}(\textbf{k}+\textbf{q},\omega^{\prime})\Bigg)\nonumber
\end{eqnarray}
The current-current correlation function can be further simplified by replacing one of the spectral functions in each term with the full $G$ or $F_{pg}$ function, resulting in the final expression
\begin{eqnarray}
\label{eq:P3_THE}\overleftrightarrow{P}(\textbf{q},i\Omega_m)&=&2\displaystyle{\sum_{\textbf{k}}}\frac{\partial\xi_{\textbf{k}-\textbf{q}/2}}{\partial\bf k}\frac{\partial\xi_{\textbf{k}-\textbf{q}/2}}{\partial\bf k}\displaystyle{\int}\frac{d\omega}{2\pi}f(\omega)\\\nonumber&&\times\Bigg(A_G(\textbf{k})G(\textbf{k}+\textbf{q},\omega+i\Omega_m)+A_G(\textbf{k}+\textbf{q},\omega)G(\textbf{k},\omega-i\Omega_m)\\\nonumber&&-A_{F_{pg}}(\textbf{k})F_{pg}(\textbf{k}+\textbf{q},\omega+i\Omega_m)-A_{F_{pg}}(\textbf{k}+\textbf{q},\omega)F_{pg}(\textbf{k},\omega-i\Omega_m)\Bigg)
\end{eqnarray}

Next we prove the transverse f-sum rule in the pseudogap state 
for this more general case of arbitrary dissipation.

Recall that the f-sum rule is expressed
\begin{eqnarray}
\displaystyle{\lim_{\textbf{q}\rightarrow0}}\displaystyle{\int}\frac{d\Omega}{\pi}\Bigg(-\frac{1}{\Omega}\textrm{Im}\overleftrightarrow{P}(\textbf{q},\Omega^+)\Bigg)=\frac{\overleftrightarrow{n}}{m}
\end{eqnarray}
where $\Omega^+=\Omega+i0^+$.

From Eq.~(\ref{eq:NUM2_THE}) we have
\begin{eqnarray}
\label{eq:KK_THE}\frac{\overleftrightarrow{n}}{m}&=&-2\displaystyle{\sum_K}\frac{\partial\xi_{\bf k}}{\partial\bf k}\frac{\partial\xi_{\bf k}}{\partial\bf k}\Bigg(G_K^2-F_{pg,K}^2\Bigg)\\\nonumber&=&-2\displaystyle{\sum_{\textbf{k}}}\frac{\partial\xi_{\bf k}}{\partial\bf k}\frac{\partial\xi_{\bf k}}{\partial\bf k}\displaystyle{\int}\frac{d\omega d\omega^{\prime}}{(2\pi)^2}\Bigg(f(\omega)-f(\omega^{\prime})\Bigg)\\\nonumber&&\times\frac{A_G(\textbf{k},\omega)A_G(\textbf{k},\omega^{\prime})-A_{F_{pg}}(\textbf{k},\omega)A_{F_{pg}}(\textbf{k},\omega^{\prime})}{\omega-\omega^{\prime}}
\end{eqnarray}

From
Eq.~(\ref{eq:32})
we have
\begin{eqnarray}	
\frac{1}{\pi\Omega}\textrm{Im}\overleftrightarrow{P}(0,\Omega^+)&=&-2\displaystyle{\sum_{\bf k}}\frac{\partial\xi_{\bf k}}{\partial\bf k}\frac{\partial\xi_{\bf k}}{\partial\bf k}\displaystyle{\int}\frac{d\omega}{2\pi}\frac{f(\omega)}{\Omega}\\\nonumber&&\times\Bigg(A_G(\textbf{k},\omega)A_G(\textbf{k},\omega+\Omega)-A_{F_{pg}}(\textbf{k},\omega)A_{F_{pg}}(\textbf{k},\omega+\Omega)\Bigg)
\end{eqnarray}

Thus
\begin{eqnarray}
\displaystyle{\int}\frac{d\Omega}{\pi}\Bigg(-\frac{1}{\Omega}\textrm{Im}\overleftrightarrow{P}(\textbf{q},\Omega)\Bigg)&=&2\displaystyle{\sum_{\bf k}}\frac{\partial\xi_{\bf k}}{\partial\bf k}\frac{\partial\xi_{\bf k}}{\partial\bf k}\displaystyle{\int}\frac{d\omega}{2\pi}\frac{d\omega^{\prime}}{\pi}\frac{f(\omega)}{\omega^{\prime}-\omega}
\\\nonumber&&\times\Bigg(A_G(\textbf{k},\omega)A_G(\textbf{k},\omega^{\prime})-A_{F_{pg}}(\textbf{k},\omega)A_{F_{pg}}(\textbf{k},\omega^{\prime})\Bigg)\\\nonumber&=&-\frac{\overleftrightarrow{n}}{m}
\end{eqnarray}
\end{widetext}

which leads to the desired result.

\bibliography{Review2.bib,Review2c.bib}
\end{document}